\documentclass[aps,prb,twocolumn,amsmath,amssymb,nofootinbib,superscriptaddress,floatfix]{revtex4}
\usepackage{graphicx,color}

\usepackage{bm} 
\usepackage{graphicx}
\usepackage{amssymb}
\usepackage{amsmath}
\usepackage{dcolumn} 

\usepackage[hypertex]{hyperref}

\begin{document}

\title{
Three-dimensional topological phase on the diamond lattice
}

\author{Shinsei Ryu}
\affiliation{
  Department of Physics,
  University of California,
  Berkeley,
  CA 94720,
  USA}

\date{\today}

\begin{abstract}
An interacting bosonic model 
of Kitaev type is proposed
on the three-dimensional diamond lattice.
Similarly to the two-dimensional Kitaev model
on the honeycomb lattice which 
exhibits both Abelian and non-Abelian phases, 
the model has two (``weak'' and ``strong'' pairing) phases.
In the weak pairing phase, 
the auxiliary Majorana hopping problem is
in a topological superconducting phase 
characterized by a non-zero winding number introduced in
A.\ P.\ Schnyder,
S.\ Ryu,
A.\ Furusaki,
and 
A.\ W.\ W.\ Ludwig,
Phys.\ Rev.\ B \textbf{78}, 195125 (2008) 
for the ensemble of Hamiltonians 
with both particle-hole and time-reversal symmetries. 
The topological character of the weak pairing phase 
is protected by a discrete symmetry.
\end{abstract}

\maketitle

\section{Introduction}

The recent discovery of $\mathbb{Z}_2$ topological insulators,
a band insulator with particular topological characters
of 
Bloch wavefunctions, 
came as a surprise.
\cite{KaneMele,Roy06,Moore06,Roy3d,Fu06_3Da,Fu06_3Db,Bernevig2006,
Koenig07,Hasan07}
On the one hand, 
$\mathbb{Z}_2$ topological insulators are
close relatives to more familiar 
integer quantum Hall (IQH) states.
\cite{Thouless82,Kohmoto}
As in an IQH state in the bulk, 
they are characterized by a topological invariant 
($\mathbb{Z}_2$ invariant).
As in an IQH state 
with boundaries,
they support stable gapless boundary states that are 
robust against perturbations. 
On the other hand, 
unlike the IQH states,
time-reversal symmetry (TRS) is a prerequisite to 
the existence of $\mathbb{Z}_2$ topological insulators.
In fact, 
as soon as the TRS of a $\mathbb{Z}_2$ topological insulator is broken,
it becomes possible to deform in a continuous manner 
a band insulator with a trivial $\mathbb{Z}_2$ topological number
into one with 
a non-trivial $\mathbb{Z}_2$ number.

Time-reversal symmetry for spin 1/2 particles 
is not the only discrete symmetry
for which a topological distinction of quantum ground
states arises.
A systematic and exhaustive 
classification of topological band insulators and 
mean-field superconductors
has been proposed in Ref.\ \onlinecite{Schnyder08}
by relying on the discrete symmetries 
of relevance to the theory of random matrices.
\cite{Zirnbauer96, Altland97}
In three spatial dimensions,
it was shown that,
besides
the $\mathbb{Z}_2$ topological insulator
in the symplectic symmetry class, 
there are precisely four more symmetry classes
in which topological insulators and/or superconductors are possible. 
\cite{KitaevKtheory,footnote}
For three out of the five symmetry classes of random matrix theory,
we introduced a topological invariant $\nu$ (winding number),
which distinguishes several different topological insulators/superconductors,
just like the Chern integer distinguishes different 
IQH states in two dimensions.
\cite{Thouless82,Kohmoto}

While the classification given in Ref.\ \onlinecite{Schnyder08} is for non-interacting fermionic 
systems, strong correlations among electrons (or spins)
might spontaneously give rise to these topological phases,
by forming a nontrivial band
structure for some, possibly emergent, fermionic excitations (e.g., spinons).
\cite{Raghu07}
It is the purpose of this paper to demonstrate how 
topological insulators (superconductors)
emerge as a result of strong correlations. 
We will show that it is possible to design an interacting bosonic model
with emergent Majorana fermion excitations,
the ground state of which is 
a topological insulator (superconductor) 
with non-vanishing winding number, $\nu\neq 0$.

Our model is a natural generalization of the spin-1/2 model on the honeycomb 
lattice introduced by Kitaev\cite{Kitaev05}
to the three-dimensional diamond lattice
with four dimensional Hilbert space per site.
The Kitaev model on the honeycomb lattice 
has two types of phases:
the so-called Abelian and non-Abelian phases.
The Abelian phase is equivalent
to the toric code model \cite{Kitaev_toric} and 
an exactly solvable model proposed by Wen, \cite{Wenbook}
which in turn is described by a $\mathbb{Z}_2$ gauge theory.
On the other hand, the non-Abelian phase
is in the universality class of the Moore-Read Pfaffian state. 
Each phase corresponds to the weak and strong pairing phases of 
two-dimensional spinless chiral $p$-wave superconductor, respectively,
\cite{Read00}
the latter of which is an example of a topological 
superconductor in symmetry class D
of Altland-Zirnbauer classification
in two dimensions.
\cite{Zirnbauer96,Altland97,Schnyder08}

Similarly to the Kitaev model on the honeycomb lattice,
the ground state of our model 
can be obtained from a Majorana fermionic ground state
(with a suitable projection procedure).
Our model has two phases, 
which we also call strong and weak pairing phases.
In particular, in the weak pairing phase, 
the ground state is 
given by a topological superconducting state in symmetry class DIII
of Altland-Zirnbauer classification,
and 
in the universality class of 
a three-dimensional analogue of the Moore-Read Pfaffian state
discussed in Ref.\ \onlinecite{Schnyder08}.
The B phase of ${ }^3\mathrm{He}$ is also in this universality class. 
\cite{Schnyder08, Roy08, Qi08}.
The topological character of the ground state is 
protected by a discrete symmetry transformation, 
which is a combination of time-reversal and 
a four-fold discrete rotation, 
the latter of which 
forms a subgroup of a continuous U(1) symmetry of our model.
Spin-1/2 models of Kitaev type on the diamond lattice,
and on other three-dimensional lattices have been constructed.
\cite{Si-Yu2007a,Si-Yu2007b,Mandal-Surendran2007}
For these models, however, 
there is no phase analogous to 
the non-Abelian phase in the original Kitaev model,
and the ground states discussed there have a vanishing winding number.
Extensions of the spin 1/2 Kitaev model to 
models with 4-dimensional Hilbert space per site have been 
studied in Refs.\ \onlinecite{Wenbook,Levin03,Hamma05,Yao08,CongjunWu08}.

\section{local Hilbert space and discrete symmetries}

We start by describing the local Hilbert space of our model, 
defined as it is at each site of some lattice.
Consider the four-dimensional Hilbert space spanned by
the orthonormal basis
\begin{align}
|\sigma \tau\rangle,
\quad
\sigma=\pm 1,
\quad
\tau = \pm 1. 
\end{align}
This space can be viewed, if we wish,
as describing 
the four-dimensional Hilbert space of
a spin 3/2 degree of freedom,
or as a direct product of two spin 1/2 Hilbert spaces. 
In the latter case, one can view these two spin 1/2 degrees of freedom
as, say,
originating from spin and orbital. 
We will denote two sets of Pauli matrices, 
$\sigma^{\mu}
=
\sigma^{0},
\sigma^{x},
\sigma^{y},
\sigma^{z}
$ 
and 
$
\tau^{\mu}
=
\tau^{0},
\tau^{x},
\tau^{y},
\tau^{z}
$
($\mu=0,1,2,3$), 
each acting on $\sigma$ and $\tau$ indices, 
with $\sigma^{0}$ and $\tau^0$
being $2\times 2$ unit matrices.

We shall represent the Hamiltonian
in terms of two sets of Dirac matrices
$\alpha^{\mu=0,1,2,3}$
(Dirac representation),
\begin{align}
\alpha^a = \sigma^a \otimes \tau^x,
\quad
\alpha^0 = \sigma^0 \otimes \tau^z =\beta=\gamma^0,
\end{align}
and
$\zeta^{\mu=0,1,2,3}$
(chiral representation),
\begin{align}
\zeta^a = -\sigma^a \otimes \tau^z,
\quad 
\zeta^0 = \sigma^0 \otimes \tau^x=\gamma_5,
\end{align}
where $a=1,2,3$.
The two sets 
$\{\alpha^{\mu}\}$ and $\{\zeta^{\mu}\}$
are related to each other by
$ 
\zeta^{\mu}=
{i}\alpha^{\mu} {i}\gamma^5 \gamma^0
=
{i}\alpha^{\mu} (\sigma^0\otimes \tau^y)
$,
and 
satisfy the Dirac algebra,
\begin{align}
\{\alpha^{\mu}, \alpha^{\nu}\} =
\{\zeta^{\mu}, \zeta^{\nu}\} = 2\delta^{\mu \nu},
\quad 
\mu,\nu=0,\ldots,3.
\end{align}

\subsection{discrete symmetries}
In the following, we will consider three 
antiunitary discrete symmetry operations,
$T$, $T'$, and $\Theta$. 
They are characterized by 
\begin{align}
T^2 = +1,
\quad
\Theta^{2}= -1.
\quad
T^{\prime 4} = -1.
\end{align}
In the sequel, we will treat two distinct
operations for time-reversal (TR). 

First, if the local Hilbert space is interpreted as
describing a spin 3/2 particle, 
the natural TR operation $\Theta$ is given by
$\Theta =\eta e^{-{i}\pi S^{3/2}_y} K$,
where $\eta$ stands for an arbitrary phase 
(will be set to one henceforth),
$S^{3/2}_y$ is a four by four matrix representing
the $y$ component of spin with $S=3/2$,
and $K$ implements the complex conjugation,
$K {i} K^{-1}= -{i}$.
If we take $|\sigma\tau\rangle$ 
to be the basis that diagonalizes $S^z$
(magnetic basis, $|3/2,m\rangle$), then
\begin{align}
\Theta = -{i}\sigma^y \otimes\tau^x K.
\end{align}
This is nothing but the charge conjugation matrix 
$C={i}\gamma^2 \gamma^0$ for the gamma matrices in
the Dirac representation.
As $\Theta$ is TRS for half-integer spin, 
$\Theta^2=-1$.
Note also that
\begin{align}
\Theta \alpha^{\mu} \Theta^{-1} = -\alpha^{\mu},
\quad
\Theta \zeta^{\mu} \Theta^{-1} = +\zeta^{\mu}.
\end{align}

Second, if the local Hilbert space is interpreted as 
describing two spin 1/2 degrees of freedom,
we can consider a TR operation $T$ defined by
\begin{align}
& T= ({i}\sigma^y)\otimes({i}\tau^y) K,
\nonumber \\
&
T \sigma^a T^{-1} = -\sigma^a,
\quad
T \tau^a T^{-1} = -\tau^a, 
\end{align}
with $a= 1,2,3$.
Note that $T^2 =+1$.
Under $T$, $\alpha$ and $\zeta$ are transformed as
\begin{align}
&
T \alpha^{\mu} T^{-1} = -\alpha_{\mu},
\quad
T \zeta^{\mu} T^{-1} = -\zeta_{\mu},
\quad
\nonumber \\
&
T {i}\gamma^5 \gamma^0 T^{-1}
=
-{i}\gamma^5 \gamma^0,
\end{align}
where covariant and contravariant vectors are
defined as
$\alpha^{\mu}=(\beta,\alpha^a)$
and
$\alpha_{\mu}=(\beta,-\alpha^a)$.

As we will see later,
while the $\sigma$-part of our Hamiltonian 
is fully anisotropic in $\sigma$ space,
the $\tau$-part of the Hamiltonian  
is invariant under a rotation around $\tau^y$ axis.
In particular, 
it is invariant under a rotation $R$
by $\pi/2$ around $\tau^y$ axis,
\begin{align}
R 
\left(
\begin{array}{c}
\tau^x \\
\tau^y \\
\tau^z \\
\end{array}
\right)
R^{-1} = 
\left(
\begin{array}{c}
\tau^z \\
\tau^y \\
-\tau^x \\
\end{array}
\right),
\,
R=
(\tau^0
+
{i}\tau^y )/
\sqrt{2}.
\end{align}
Under $R$, $\alpha$ and $\zeta$ are transformed as
\begin{align}
&
R\alpha^{\mu} R^{-1} = -\zeta^{\mu},
\quad
R\zeta^{\mu} R^{-1} = +\alpha^{\mu},
\nonumber \\
&
R {i}\gamma^5 \gamma^0
R^{-1} = +{i}\gamma^5 \gamma^0.
\end{align}
By combining $T$ with $R$
we can define yet another antiunitary operation, $T'=RT$,
\begin{align}
&
T' 
= RT
=
({i}\tau^y -\tau^0)
{i}\sigma^y
K/\sqrt{2},
\nonumber \\
&
T' \sigma^a T^{\prime -1} = -\sigma^a,
\quad
T'
\left(
\begin{array}{c}
\tau^x \\
\tau^y \\
\tau^z \\
\end{array}
\right)
T^{\prime -1}
=
\left(
\begin{array}{c}
-\tau^z \\
-\tau^y \\
+\tau^x \\
\end{array}
\right).
\end{align}
Below, 
with a slight abuse of language, 
we will call this operation $T'$
time-reversal operation (TR). 
When applied to $\alpha$ and $\zeta$,
\begin{align}
&
T' \alpha^{\mu}T^{\prime -1}
=
+\zeta^{\ }_{\mu},
\quad
T'\zeta^{\mu}T^{\prime -1}
=
-\alpha^{\mu},
\nonumber \\
&
T'
{i}\gamma^5 \gamma^0
T^{\prime -1}
=
-{i}\gamma^5 \gamma^0,
\label{eq: TRS on alpha and zeta}
\end{align}
i.e.,
TRS $T'$ exchanges $\alpha$ and $\zeta$,
and covariant and contravariant vectors.
Notice that 
\begin{align}
T^{\prime 2} 
&=
{i}\tau^y,
\quad
T^{\prime 4} =
-1.
\end{align}

\section{Hamiltonian}

In the two-dimensional Kitaev model,
different types of interactions 
(represented by three $2\times 2$ Pauli matrices)
are assigned 
to three distinct types of bonds 
that are defined by their different orientation
in the honeycomb lattice.
It is this anisotropic nature of the interaction
that makes the Kitaev model exactly solvable.
This construction can be extended 
to the diamond lattice where there are 
four distinct types of bonds with different orientation.

\begin{figure}[t]
\centering
\includegraphics[width=0.35\textwidth]{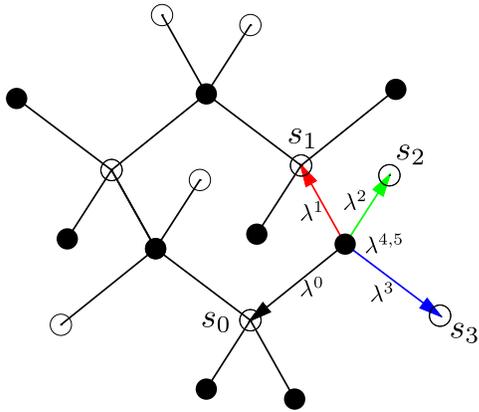}
\caption{
(Color online) 
The diamond lattice and the six Majorana fermions
$\lambda^{0,\ldots, 5}$.
Sites on the sublattice $A$ ($B$) are denoted by 
an open (filled) circle.
        } 
\label{fig:diamond_lattice}
\end{figure}

The diamond lattice is bipartite and 
consists of two interpenetrating fcc lattices shifted by
$a(-1,1,-1)/4$ along the body diagonal,
where $a$ is the lattice constant. 
We label sites $r_A$ and $r_B$
on the two sublattices $A$ and $B$ of the diamond lattice as
\begin{align}
r_A = 
\sum_{i=1}^3 m_i {a}_i ,
\quad
r_B = {r}_A +{s}_0,
\quad 
m_i\in\mathbb{Z},
\end{align}
where the primitive vectors ${a}_i$ are given by
\begin{eqnarray}
{a}_1
=
\frac{a}{2}
\left(
\begin{array}{c}
1 \\
1 \\
0
\end{array}
\right),
\quad
{a}_2
=
\frac{a}{2}
\left(
\begin{array}{c}
0 \\
1 \\
1
\end{array}
\right),
\quad
{a}_3
=
\frac{a}{2}
\left(
\begin{array}{c}
1 \\
0 \\
1
\end{array}
\right).
\end{eqnarray}
We have also introduced 
the 3-component vectors
\begin{eqnarray}
{s}_1
=
\frac{a}{4}
\left(
\begin{array}{c}
1 \\
1 \\
1
\end{array}
\right),
\quad
{s}_2
=
\frac{a}{4}
\left(
\begin{array}{c}
-1 \\
-1 \\
1
\end{array}
\right),
\nonumber \\
{s}_3
=
\frac{a}{4}
\left(
\begin{array}{c}
1 \\
-1 \\
-1
\end{array}
\right),
\quad
{s}_0
=
\frac{a}{4}
\left(
\begin{array}{c}
-1 \\
1 \\
-1
\end{array}
\right),
\end{eqnarray}
which connect
the nearest neighbor sites.
(Fig.\ \ref{fig:diamond_lattice}).

The Hamiltonian we study in this paper is 
defined by 
\begin{align}
H 
&=
-\sum^3_{\mu=0} J_{\mu}
\sum_{\mu-\mathrm{links}} 
\left(
\alpha^{\mu}_j \alpha^{\mu}_k
+
\zeta^{\mu}_j \zeta^{\mu}_k
\right).
\label{eq: Hamiltonian}
\end{align}
Here, 
the sites $j$ and $k$ are end points of a link
of type $\mu$.
There are 
four types of links $\mu=0,1,2,3$ in the diamond lattice 
which can be 
distinguished by their orientations.
Hamiltonian (\ref{eq: Hamiltonian}) can also be written in terms of $\sigma^{\mu}$ and $\tau^{\mu}$ as
\begin{align}
H 
&=
-\sum^3_{\mu=0} J_{\mu}
\sum_{\mu-\mathrm{links}} 
\sigma^{\mu}_j 
\sigma^{\mu}_k 
\left(
\tau^x_j \tau^x_k
+
\tau^z_j \tau^z_k
\right).
\label{eq: hamiltonian}
\end{align}
This Hamiltonian is invariant under
discrete symmetries,
$T$, $R$, $T'$ and $\Theta$,
and enjoys a U(1) symmetry for rotation
around $\tau^y$ axis.

\section{Majorana fermion representation}

Let us consider the (local) Hilbert space in which we have
six Majorana fermions $\{\lambda^p \}_{p=0,\ldots,5}$
per site, which satisfy\cite{Levin03,Wenbook}
\begin{align}
\left\{\lambda^p, \lambda^q\right\} = 2\delta^{pq},
\quad
p,q=0,\ldots,5.
\end{align}
To construct 4-dimensional Hilbert space
out of the 8-dimensional Hilbert space, 
we introduce the fermion number operator by 
\begin{align}
D &: = {i} \prod_{p=0}^5 \lambda^p,
\quad
D^2 = 1.
\end{align}
The eigenvalue of $D=\pm 1$ can then be used to 
select a 4-dimensional subspace of the full Hilbert space,
which will be called the physical subspace.

We can construct 15 generators of $\mathrm{so}(6)$
from the Majorana fermions as
\begin{align}
&
\Gamma^{pq}=
{i}
\lambda^p \lambda^q,
\quad p\neq q.
\end{align}
Within the physical subspace, 
$\alpha^{\mu}$ can be expressed as
\begin{align}
\alpha^{\mu}= \Gamma^{\mu 4},
\quad
\mu=0,1,2,3.
\end{align}
Since $\Gamma^{45}$ anti-commutes with $\alpha^{\mu}$, 
we make the identification 
\begin{align}
\Gamma^{45}
=
\alpha^{1}
\alpha^{2}
\alpha^{3}
\alpha^{0}
=
\sigma^0 \otimes \tau^y
=
{i}\gamma^5 \gamma^0.
\end{align}
The second set of the gamma matrices 
$\{\zeta^{\mu}\}$ is
\begin{align}
\zeta^{\mu} = {i}\alpha^{\mu}\Gamma^{45}
=
\Gamma^{\mu 5}.
\end{align}

The Majorana fermions naturally inherit
the symmetry operations on $\alpha$ and $\zeta$.
The symmetry conditions are automatically satisfied if we define
$T$,$R$, and $T'$ operations on Majorana fermions by
\begin{align}
&
T \lambda^{\mu}T^{-1}
=
\lambda_{\mu},
\quad
T \lambda^{s}T^{ -1}
=
\lambda^{s},
\quad
s=4,5,
\nonumber \\
&
R \lambda^{\mu}R^{-1}
=
-\lambda^{\mu},
\quad
R 
\left(
\begin{array}{cc}
\lambda^{4} \\
\lambda^{5} \\
\end{array}
\right)
R^{-1}
=
{i}s^y
\left(
\begin{array}{cc}
\lambda^{4} \\
\lambda^{5} \\
\end{array}
\right),
\nonumber \\
&
T' \lambda^{\mu}T^{\prime -1}
=
-\lambda_{\mu},
\quad
T' 
\left(
\begin{array}{c}
\lambda^4 \\
\lambda^5 
\end{array}
\right)
T^{\prime -1}
=
{i}s^y
\left(
\begin{array}{c}
\lambda^4 \\
\lambda^5 
\end{array}
\right).
\end{align}
Here, the definitions of 
the covariant ($\lambda^{\mu}$)
and contravariant ($\lambda^{\mu}$) vectors follow
from those of $\alpha^{\mu}$ ($\zeta^{\mu}$) and 
$\alpha_{\mu}$ ($\zeta_{\mu}$),
and
we have introduced another
set of Pauli matrices $s^{\mu=0,1,2,3}$ 
acting on $\lambda^{4,5}$
with $s^0$ being $2\times 2$ unit matrix.
The discrete rotation operator 
$R$ can be written in terms of the Majorana fermions $\lambda$ as
$e^{\mathrm{i}\pi \lambda^4 \lambda^5 /4}.$

\section{solution through a Majorana hopping problem}

In terms of the Majorana fermions
the Hamiltonian can be written as
\begin{align}
H 
&=
{i} \sum^3_{\mu=0}
J_{\mu} \sum_{\mu-\mathrm{links}} 
u_{jk}
\left(
\lambda^4_j \lambda^4_k
+
\lambda^5_j \lambda^5_k
\right).
\label{eq: Majorana, nn}
\end{align}
where we have introduced a link operator by
\begin{align}
u_{jk}:={i}\lambda^{\mu_{jk}}_j \lambda^{\mu_{jk}}_k,
\end{align}
with $\mu_{jk}=0,1,2,3,4$ depending on the orientation of the
link ending at sites $j$ and $k$.
Note that TR ($T$ or $T'$) operation flips the sign of a link operator,
\begin{align}
&
T u_{jk} T^{-1} = -u_{jk},
\quad
R u_{jk} R^{-1} = +u_{jk},
\nonumber \\
&
T' u_{jk} T^{\prime -1} = -u_{jk}.
\end{align}
(This is also the case for TRS on 
the link operators in the spin-1/2 honeycomb lattice Kitaev model.)
When necessary, this sign flip can be removed
by a subsequent gauge transformation for 
Majorana fermions $\lambda^{4,5}$ on either one of sublattices, if 
the underlying lattice structure is bipartite (see below).

What is essential to observe is that
all $u_{jk}$ appearing in the Hamiltonian
commute with each other and with the Hamiltonian. 
They can thus be replaced by their eigenvalues
$u_{jk}=\pm 1$,
and the interacting Hamiltonian reduces to,
for a fixed configuration of the $\mathbb{Z}_2$ gauge
field $\{u_{jk}\}$, 
a simple hopping model of Majorana fermions. 
Observe that both $\lambda^4$ and $\lambda^5$ Majorana
fermions feel the same $\mathbb{Z}_2$ gauge field. 
The ground state of the Hamiltonian can then be obtained by
first picking up the $\mathbb{Z}_2$ gauge field configuration
that gives the lowest ground state energy for 
the Majorana hopping problem, and then projecting 
the resulting fermionic ground state onto the physical Hilbert space. 
According to Lieb's theorem, \cite{Lieb94}
the $\mathbb{Z}_2$ gauge field configuration
that gives the lowest ground state energy 
has zero $\mathbb{Z}_2$ vortex for all hexagons,
and hence we can take $u_{jk}=1$ for all links. 

For notational convenience, 
for a Majorana fermion at the $j$th site located at $r_{A}$ ($r_{B}$) on the
sublattice $A$ ($B$),
we denote $a^s_{r_A} :=\lambda^s_j$
($b^s_{r_B} :=\lambda^s_j$). 
With periodic boundary condition and with the Fourier transformation, 
$
a^s_{r_A}
=
\sum_{k}
e^{{i}k\cdot r_A} a^s_k
/\sqrt{|\Lambda_A|}
$
and
$
b^s_{r_B}
=
\sum_{k}
e^{{i}k\cdot r_B} b^s_k
/\sqrt{|\Lambda_B|},
$
where $|\Lambda_{A,B}|$ is the total number of sites
on the sublattice $A,B$, respectively, 
the Majorana hopping Hamiltonian in the momentum space is 
\begin{align}
H_{\mathrm{MH} }
=
\sum_{s}
\sum_{k} 
\left(
\begin{array}{cc}
a^s_{-k} , & b^s_{-k}
\end{array}
\right)
\mathcal{H}(k)
\left(
\begin{array}{c}
a^s_{k} \\
b^s_{k}
\end{array}
\right),
\label{eq: Majorana H in k}
\end{align}
where we have defined
\begin{align}
\mathcal{H}(k):=
\left(
\begin{array}{cc}
& {i}\Phi(k)\\
-{i}\Phi^*(k)
&
\end{array}
\right),
\quad
\Phi(k):=\sum_{\mu=0}^{3} J_{\mu} e^{{i}k\cdot s_{\mu}},
\end{align}
and noted
$a^{\ }_{-k}=a^{\dag}_{k}$ when $k\neq 0$.
The energy spectrum $E(k)$ is
given by 
$E(k)=
\pm
\sqrt{
|\Phi(k)|^2
}$,
with two-fold degenerate for each $k$.

\subsection{symmetries and topology of the Majorana hopping Hamiltonian}

We have reduced the interacting bosonic model to 
the Majorana hopping problem.
This auxiliary Majorana hopping Hamiltonian is,
in the terminology of Altland and Zirnbauer, 
in symmetry class D, 
\cite{Zirnbauer96,Altland97}
i.e., the ensemble of quadratic Hamiltonians
describing Majorana fermions.
(See Appendix \ref{appendix A}.)
In more general situations (which we will consider below), 
the auxiliary Majorana hopping Hamiltonian is given by
\begin{align}
H_{\mathrm{MH}} 
&=
\sum_{k}
\left(
\begin{array}{cccc}
a^4_{-k} , & a^5_{-k} , & b^4_{-k}, & b^5_{-k}
\end{array}
\right)
\mathcal{X}(k)
\left(
\begin{array}{c}
a^4_{k} \\
a^5_{k} \\
b^4_{k} \\
b^5_{k}
\end{array}
\right),
\end{align}
where $\mathcal{X}$ describes
a Hamiltonian for Majorana fermions,
and satisfies
\begin{align}
\mathcal{X}^{\dag}(k)= \mathcal{X}(k),
\quad
\mathcal{X}^T(-k) = -\mathcal{X}(k).
\label{eq: class D}
\end{align}
This is the defining property of symmetry class D. 
Below, to describe the $4\times 4$ structure of 
the single-particle Majorana hopping 
Hamiltonian $\mathcal{X}(k)$,
we introduce yet another
set of Pauli matrices $c^{\mu=0,1,2,3}$ acting on sublattice indices.

If our bosonic model further satisfies TRS $T'$, 
the Hamiltonian $\mathcal{X}$ for
the  auxiliary Majorana hopping problem respects 
\begin{align}
c^z ({i}s^y) \mathcal{X}^T(-k) (-{i}s^y)c^z
= \mathcal{X}(k),
\label{eq: class DIII}
\end{align}
where the factor $c^z$ can be thought of as a gauge transformation,
adding a phase factor $e^{{i}\pi}$ 
for Majorana fermions on $B$ sublattice $b^s$,
and can be removed by a unitary transformation $b^s\to -b^s$.
With this further condition arising from $T'$,
the relevant Altland-Zirnbauer 
symmetry class is class DIII. 
(See Appendix \ref{appendix A}.)

In Ref.\ \onlinecite{Schnyder08}, it has been shown that 
the space of all possible quantum ground states in class DIII
in three spatial dimensions is partitioned into 
different topological sectors, each labeled by an integer topological invariant
$\nu$.
To uncover this topological structure and
introduce the winding number,
we observe that all Hamiltonians in symmetry class DIII can be
brought into a block off-diagonal form.
For $\mathcal{X}$, this is done by a unitary transformation
\begin{align}
U &= U^{\ }_2 U^{\ }_1 ,
\quad 
\mathcal{X} \to
\tilde{\mathcal{X}}
=
U^{\ }_2 U^{\ }_1 \mathcal{X} U^{\dag}_1 U^{\dag}_2, 
\end{align}
where the first unitary transformation
rotates $s^y\to U_1 s^y U^{\dag}_1=-s^z$,
\begin{align}
U_1 &=
(s^0 -{i}s^x)/\sqrt{2}, 
\end{align}
whereas the second unitary transformation
exchanges 2nd and 4th entries, 
\begin{align}
U_2 &=
(s^0 + s^z) c^0/2
+ (s^0 - s^z)c^x/2.
\end{align}
The combination of $U_1$ and $U_2$
diagonalizes $s^y c^z$ as
\begin{align}
U^{\ }_2 U^{\ }_1 s^y c^z U^{\dag}_1 U^{\dag}_2 
&=
-\mathrm{diag}\,\left(1,1,-1,-1\right).
\end{align}
After the unitary transformation, we find
\begin{align}
\tilde{\mathcal{X}}(k)
&=
\left(
\begin{array}{cc}
0 & D(k) \\
D^{\dag}(k) & 0
\end{array}
\right).
\end{align}
This block off-diagonal structure is inherited 
to the spectral projector $P(k)$,
$P^2 = P$, 
which projects onto the space of filled Bloch states at each $k$,
\begin{eqnarray}
2P(k)-1
&=
\left(
\begin{array}{cc}
0 & q(k) \\
q^{\dag}(k) & 0
\end{array}
\right),
\quad
q^{\dag} q=1.
\end{eqnarray}
The integer topological invariant 
is then defined,
from the off-diagonal block of the projector, as\cite{Schnyder08}
\begin{equation}
\nu[q]
=
\int_{\mathrm{Bz}}
\frac{d^3 k\,}{24\pi^2}
\epsilon^{\mu\nu\rho}\,
\mathrm{tr}\left[
(q^{-1}\partial_{\mu}q) \cdot
(q^{-1}\partial_{\nu}q) \cdot
(q^{-1}\partial_{\rho}q)
\right],
\label{eq: winding number}
\end{equation}
where $\mu,\nu,\rho=k_{x}, k_{y}, k_{z}$,
and the integral extends over the first 
Brillouin zone (Bz).
\cite{SalomaaVolovik88}
The non-zero value of the winding number signals 
a non-trivial topological structure, an observable consequence
of which is the appearance of gapless surface Majorana fermion modes.

\section{strong  pairing phase}

When one of the coupling $J_{\mu}$ is strong enough
compared to the others, the spectrum for the Majorana fermions
is gapped. In this phase, the winding number $\nu$
is zero. This phase can be called ``strong pairing phase'',
following the similar phase in the BCS pairing model. 
Because of the trivial winding number, there is no surface
stable fermion mode in the auxiliary Majorana hopping Hamiltonian,
when it is terminated by a surface. 
The properties of this phase can be studied by
taking the limit $J_{0}\gg J_{a=1,2,3}>0$, say, 
and then developing a degenerate perturbation theory.
In this limit of isolated links, 
there are four degenerate ground states for each link
ending at the two sites $(r_A, r_B)=(r_A,r_A+s_0)$,
\begin{align}
\frac{1}{\sqrt{2}}
\left(
|\uparrow\rangle^{\tau}_{r_A}
|\uparrow \rangle^{\tau}_{r_B} + 
|\downarrow\rangle^{\tau}_{r_A}
|\downarrow \rangle^{\tau}_{r_B}
\right)
|\sigma_{r_A} \rangle^{\sigma}_{r_A}
|\sigma_{r_B}\rangle^{\sigma}_{r_B}
\end{align}
with $\sigma_{r_{A,B}} =\pm 1$
where $|\cdots \rangle^{\sigma,\tau}_r$
represents the state for $\sigma_{r}$, and
$\tau_{r}$, respectively.
The effective Hamiltonian acting on
these degenerate ground states is defined on
the cubic lattice, since each $J_0$ link at $(r_A,r_A+s_0)$
is connected to six neighboring links
at $r_A\pm a_i$, where $a_i = s_0+s_i$ with $i=1,2,3$,
upto the 4th order in the degenerate perturbation theory.
If we use notations,
$\sigma^{\mu}_{A}\to \rho^{\mu}$,
$\sigma^{\mu}_{B}\to \mu^{\mu}$,
the effective Hamiltonian upto the 4th order
in the degenerate perturbation theory is (upto constant terms)
\begin{align}
H_{\mathrm{eff}} 
&=
\frac{5 }{64 }
\sum_{
\substack{
(i,j,k)=
(x,y,z),\\
(y,z,x),(z,x,y)
}
}
\frac{J^2_i J^2_j}{J^3_0}
\sum_p F_p,
\end{align}
where $p$ stands for a plaquette
surrounded by $r$, $r+a_{i}$ $r+a_{j}$ and $r+a_i+a_j$,
and 
\begin{align}
F_p
=
\big(
\rho^k
\mu^0
\big)_{r}
\big(
\rho^j
\mu^i
\big)_{r+a_i}
\big(
\rho^i
\mu^j
\big)_{r+a_j}
\big(
\rho^0
\mu^k
\big)_{r+a_i+a_j}.
\end{align}
A similar model on the cubic lattice was discussed
in Refs.\ \onlinecite{Levin03, Wenbook, Hamma05}.

\section{weak pairing phase}

\begin{figure}[t]
\includegraphics[width=0.2\textwidth]{diamond_nnn_tex_2.epsi}
\includegraphics[width=0.2\textwidth]{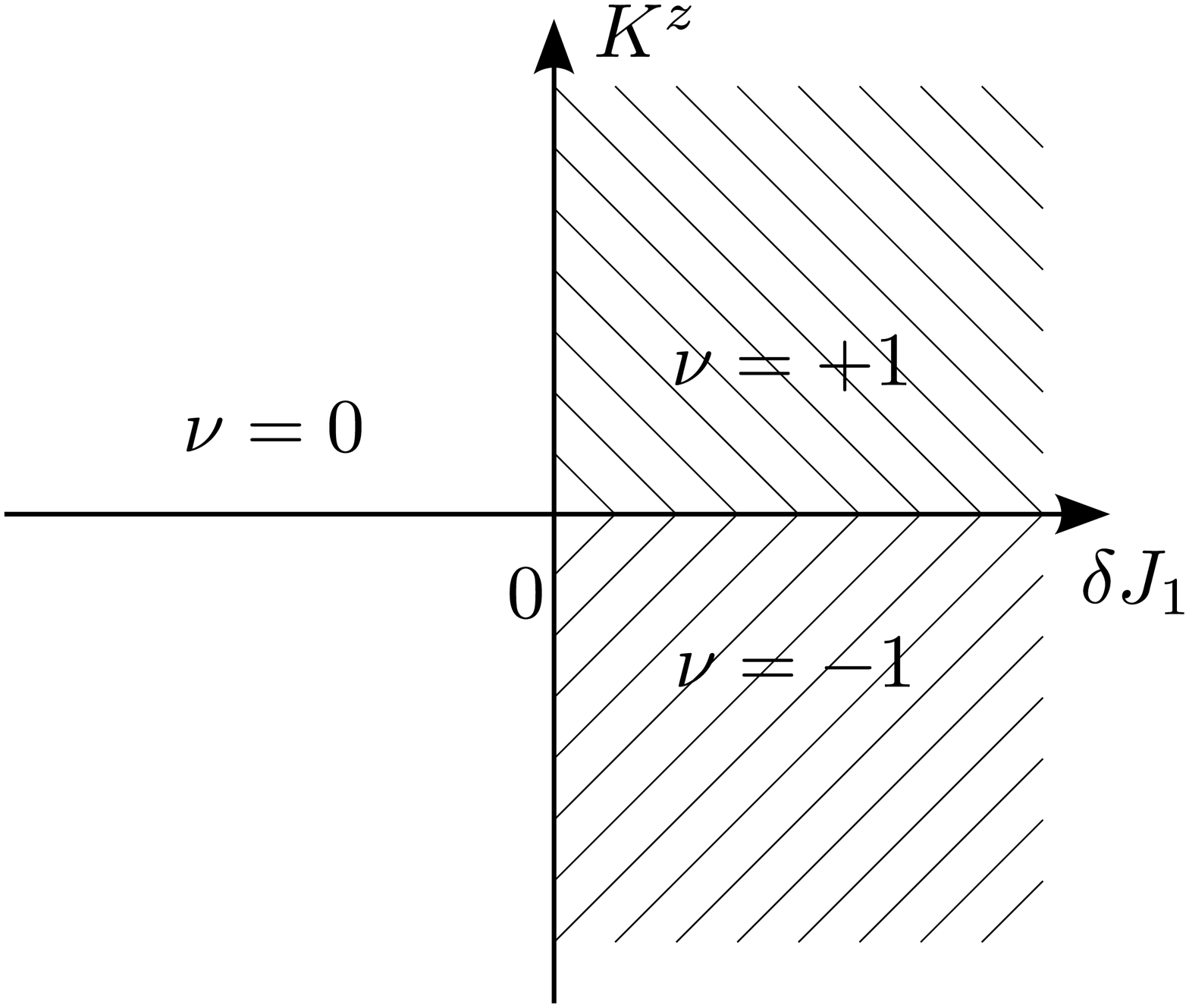} \\
\includegraphics[width=0.23\textwidth]{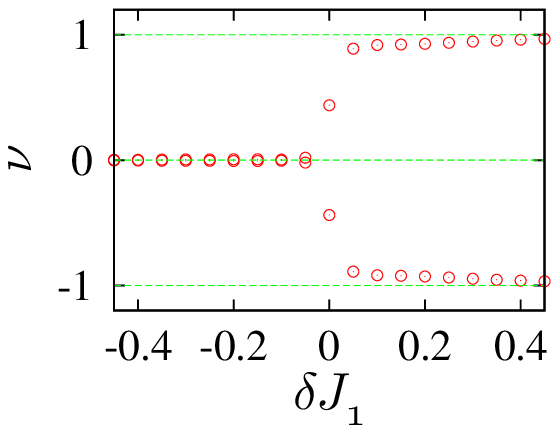}
\includegraphics[width=0.23\textwidth]{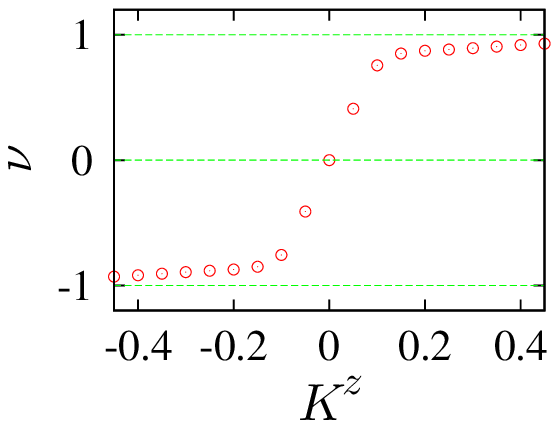} 
\caption{
\label{fig: winding number}
(Color online) 
(Top left)
The choice of the second nearest neighbor couplings
$K^x_{jlk}$ (blue links) and $K^z_{jlk}$  (red links).
(Top right) 
The phase diagram in term of 
$K^z$ and $\delta J_1$ with $J_{2,3,4}=J=2$, and $K^x=1$. 
(Bottom) 
The numerical evaluation 
of the winding number as a function of 
the second neighbor coupling $K^z$
and the distortion $\delta J_1$.
In the left panel, 
the winding number is computed for
$K^z=\pm 1$ with changing $\delta J_1$ continuously,
whereas in the right panel $\delta J_1$ is fixed, $\delta J_1 = 1$.
For $K^z = 1$ ($K^z=-1$),
$\nu=1$ ($\nu=-1$)
when $\delta J_1 >0$. 
}
\end{figure}

When all $J_{\mu}$ are equal,
$J_{\mu}=J$,
the energy spectrum of the Majorana hopping Hamiltonian
Eq.\ (\ref{eq: Majorana H in k})
has lines of zeros (line nodes) in momentum space. 
This gapless nature is, however, not stable against 
perturbations that respect TRS $T'$.
This can be illustrated by 
taking a four ``spin'' perturbation 
in the gapless phase,
defined on three sites $j,k,l$,
where sites $j$ and $k$ are 
two different nearest neighbors of site $l$.
Let us take, as an example, 
\begin{align}
\alpha^0_j
({i}\alpha^0 \alpha^1)_l
\alpha^1_k
&=
\alpha^0_j
\alpha^2_l \zeta^3_l
\alpha^1_k
=
-
\alpha^0_j
\alpha^3_l \zeta^2_l
\alpha^1_k
\nonumber \\
&=
\Gamma^{04}_j
\Gamma^{24}_l
\Gamma^{35}_l
\Gamma^{14}_k
\nonumber \\
&=
{i}
u_{jl}\lambda^4_j
\times
D_l
\times
u_{lk}
 \lambda^4_k,
\label{eq: 4 spin half}
\end{align}
where we take the link emanating from sites $j$ ($k$) and $l$ to be 
parallel to $s_0$ ($s_1$).
If perturbations of this type are small enough,
relative to the excitation energy of 
a $\mathbb{Z}_2$ vortex loop (line),
i.e., an excitation which flips the sign of $\mathbb{Z}_2$ flux
threading hexagons, 
we can contain ourselves in the vortex-free sector
where $u_{jk}=+1$.
Thus, the above four ``spin'' perturbation
leads to a next nearest neighbor hopping terms 
of the Majorana fermions.
To respect TRS $T'$,
Eq.\ (\ref{eq: 4 spin half}) can be supplemented 
with its TRS partner,
$\zeta ^0_j \alpha^3_l \zeta^2_l \zeta^1_k=
T' \alpha^0_j \zeta^3_l \alpha^2_l \alpha^1_k
T^{\prime -1}$,
leading to a perturbation
\begin{align}
H^z_{\mathrm{nnn}}
&=
\sum_{\langle\!\langle jlk\rangle\! \rangle}
K^z_{jlk}
 \left[
 {i} \alpha^{\mu}_j \alpha^{\mu}_l \alpha^{\nu}_l \alpha^{\nu}_k 
 -
 \left(
 \alpha \leftrightarrow \zeta
 \right)
 \right]
\nonumber \\
&=
{i} 
\sum_{\langle\!\langle jlk\rangle\! \rangle}
K^z_{jlk}
u_{jl}u_{lk}
\left(
\lambda^4_j \lambda^4_k
-
\lambda^5_j \lambda^5_k
\right),
\label{eq: nnn z}
\end{align}
where the summation extends
over all sites labeled by $l$ and their nearest
neighbors $j$ and $k$, 
with 
the link emanating from sites $j$ ($k$) and $l$ 
parallel to $s_{\mu}$ ($s_{\nu}$),
and 
$K^z_{jlk}\in \mathbb{R}$. 
Similarly, the following perturbation 
defined on three sites $j,l,k$
is also allowed by TRS $T'$,
\begin{align}
H^x_{\mathrm{nnn}}
&=
\sum_{\langle\!\langle jlk\rangle\! \rangle}
K^x_{jlk}
\left[
 \alpha^{\mu}_j 
(\alpha^{\mu} {i} \gamma^5 \gamma^0 \zeta^{\nu})_l 
\zeta^{\nu}_k 
+
(\alpha\leftrightarrow \zeta)
\right]
\nonumber \\
&=
{i} 
\sum_{\langle\!\langle jlk\rangle\! \rangle}
K^x_{jlk}
u_{jl}u_{lk}
\left(
\lambda^4_j \lambda^5_k
+
 \lambda^5_j \lambda^4_k
\right)
\label{eq: nnn x}
\end{align}
with $K^x_{jlk}\in \mathbb{R}$. 
We choose $K^{x,z}_{jlk}$ in such a way that these perturbations
lead to the following next nearest neighbor terms in 
the Majorana hopping Hamiltonian
(see Fig.\ \ref{fig: winding number})
\begin{align}
H^z_{\mathrm{nnn}}
=&
\frac{K^z}{2 {i}}
\sum_r 
\lambda^{T}_r s^z
\lambda_{r+s_1-s_3}
+
\mathrm{h.c.},
\nonumber \\
H^x_{\mathrm{nnn}}
=&
\frac{K^x }{2 {i}}
\sum_r 
\lambda^{T}_r s^x 
\Big[
\lambda_{r+s_0-s_2}
+
\lambda_{r+s_2-s_3}
+
\lambda_{r+s_3-s_0}
\Big]
\nonumber \\
&
+
\mathrm{h.c.},
\label{eq: Majorana nnn}
\end{align}
where $\lambda^T=(\lambda^4,\lambda^5)$
and $K^{x,z}\in \mathbb{R}$.

With these perturbations, 
the Majorana hopping Hamiltonian in momentum space is given by
\begin{align}
\mathcal{X}(k)&=
\left(
\begin{array}{cc}
\Theta(k) & {i}\Phi(k)\\
-{i}\Phi^*(k) & -\Theta(k)
\end{array}
\right),
\end{align}
where $\Phi(k)$ comes from the 
nearest neighbor hopping (\ref{eq: Majorana, nn}), 
whereas the off-diagonal part $\Theta(k)$ comes from 
the next nearest neighbor hopping terms (\ref{eq: Majorana nnn})
and is given by
\begin{align}
\Theta(k) &=
\Theta^x(k)s^x
+
\Theta^z(k)s^z,
\nonumber \\
\Theta^x(k)
&=
K^x
\left[
\sin \frac{k_x-k_y}{2}
+
\sin \frac{k_y-k_z}{2}
+
\sin \frac{k_z-k_x}{2}
\right],
\nonumber \\
\Theta^z(k)
&= 
K^z 
\sin
\frac{k_y+k_z}{2}.
\end{align}
Observe that this Hamiltonian
indeed satisfies class DIII conditions 
(\ref{eq: class D}) and (\ref{eq: class DIII}).
It can then be made block off-diagonal
by the unitary transformation $U$, with the
off-diagonal block being given by
\begin{align}
D(k) &=
-\mathrm{Im}\,\Phi(k) s^0
+\Theta^z(k) {i} s^x
\nonumber \\
&
\qquad 
+\Theta^x(k) {i} s^y
+\mathrm{Re}\,\Phi(k) {i} s^z.
\end{align}
The energy spectrum at momentum $k$ is given by
$E(k)=\pm \sqrt{
|\Phi(k)|^2 + 
[\Theta^x(k)]^2
+
[\Theta^z(k)]^2
}
$.

With these perturbations, we can indeed
lift the degeneracy except at three points $k=Q_{x,y,z}$
in the Bz, where
\begin{align}
Q_x = 
\frac{2\pi}{a}
\left(
\begin{array}{c}
1 \\
0 \\
0 \\
\end{array}
\right),
\,\,
Q_y = 
\frac{2\pi}{a}
\left(
\begin{array}{c}
0 \\
1 \\
0 \\
\end{array}
\right),
\,\,
Q_z = 
\frac{2\pi}{a}
\left(
\begin{array}{c}
0 \\
0 \\
1 \\
\end{array}
\right).
\end{align}
The dispersion around these points is Dirac-like,
\begin{eqnarray}
\mathcal{X}(Q_a+q)
\!\!&\sim&\!\!
 J q_a c^y
 + K^x (q_b-q_c) s^x c^z
 \nonumber \\
 &&
 + \frac{K^z}{2} (q_y+q_z) s^z c^z,
 \end{eqnarray}
where $(a,b,c)$ is a cyclic permutation of $(x,y,z)$. 
These three-dimensional Dirac fermions
can be made massive by further adding 
a slight distortion in the nearest
neighbor hopping, $J_1 \to J_1 +\delta J_1$, say.
This gives rise to, at $Q_{x,y,z}$, 
a perturbation to $\mathcal{X}(k)$
which takes the form of a mass term to Dirac fermions,
$-\delta J_1 c^x$.

For definiteness, we now set $J_{\mu=2,3,4}=J=2$,
$K^x=1$, 
$J_1=J+\delta J_1$,
and vary $K^z$ and $\delta J_1$.
In the $(\delta J_1, K^z)$-plane,
there are phase boundaries represented by
$\delta J_1=0$ 
and the half-line $K^z=0$ with  $\delta J_1 >0$
(Fig.\ \ref{fig: winding number}).
On the line $\delta J_1=0$, the spectrum is Dirac like
except at the origin $(\delta J_1, K^z)=(0,0)$
where the band gap closes at $Q_{x,y,z}$
quadratically in one direction in momentum space
(a similar gapless point is discussed in Ref.\ \onlinecite{CongjunWu08}).
To determine the topological nature of 
the three gapped phases,
the first and second quadrants in $(\delta J_1, K^z)$-plane,
and the region $\delta J_1 <0$, 
we computed the winding number, by 
numerically integrating the formula Eq.\ (\ref{eq: winding number}).
Integral (\ref{eq: winding number}) quickly converges,  
to a quantized value $\nu=0,\pm 1$
as we increase the number of mesh in momentum space.
While the winding number is identically zero when $\delta J_1<0$, 
it takes either $\nu=+1$ or $\nu=-1$ in the phases $\delta J_1>0$,
depending on the sign of $K^z$. 
The complete structure of the phase diagram 
including the value of the winding number
is presented in Fig.\ \ref{fig: winding number}.
In the phases with non-zero winding number, 
there appears a gapless and stable surface Majorana fermion mode,
when the Majorana hopping Hamiltonian is truncated by a boundary,
signaling non-trivial topological character in the bulk.

\section{discussion}

We have constructed a three-dimensional interacting bosonic model
which exhibits a topological band structure
for emergent Majorana fermions. 
We thus take a first step to explore
topological superconductors
arising from interactions, rather than 
given by some external parameters at the single particle level,
such as external magnetic field or spin orbit coupling. 
Although the Kitaev model does not look
particularly realistic 
as it is anisotropic both in real and spin spaces, 
it has played an important role in deepening our understanding 
on two-dimensional topological order. 
(See, for example, 
Refs.\ 
\onlinecite{
Feng_Zhang_Xiang2006,
Baskaran2006,
DHLee2007,
HanDongChen_Nussinov07,
HanDongChen_JiangpingHu07,
YueYu2007,
Yao07,
Schmidt2008}).
Also, there have been a proposal to 
realize the Kitaev model in terms of 
cold polar molecules on optical lattices
\cite{Duan2003,Micheli2006}
and superconducting quantum circuits.
\cite{Nori08}
Interactions 
which are anisotropic both in real and internal spaces
can appear in systems with orbital degrees of freedom, 
such as the orbital compass model.
Indeed, it is worth emphasizing that our model,
in the absence of four spin interactions,
possesses a U(1) rotation symmetry
unlike the original Kitaev model and its variants. 
Thus, say,
identifying $\tau$ as a spin 1/2 degree of freedom
and $\sigma$ as an orbital degree of freedom, 
it might be realized as a XY analogue of the Kugel-Khomskii model. 
Finally, while our model is designed to have 
a Gutzwiller-type projected wavefunction
as its exact ground state,
such ground state wavefunctions can appear 
in much wider context, 
which can be explored, e.g., in terms of a variational
approach with slave particle mean field theories.
\cite{Wang08}

\acknowledgments

The author acknowledges helpful interactions with
Akira Furusaki,
Andreas Ludwig, 
Christopher Mudry, 
Andreas Schnyder, 
Ashvin Vishwanath,
Grigory Volovik,
and Congjun Wu.
This work has been supported by Center for Condensed Matter Theory
at University of California, Berkeley.

\appendix

\section{class DIII symmetry class}
\label{appendix A}

In this appendix, we review 
the symmetry classification of the
Bogoliubov-de Gennes (BdG) Hamiltonians 
by Altland and Zirnbauer,\cite{Altland97}
which is relevant to 
our auxiliary Majorana hopping problems.
We consider the following general form 
of a BdG Hamiltonian for the dynamics of 
fermionic quasiparticles deep inside the superconducting
state of a superconductor
\begin{eqnarray}
H
=\frac{1}{2}
\left(
\begin{array}{cc}
\boldsymbol{c}^{\dag}, & \boldsymbol{c}
\end{array}
\right)
\mathcal{H}
\left(
\begin{array}{c}
\boldsymbol{c}^{\ } \\
\boldsymbol{c}^{\dag}
\end{array}
\right),\,\,
\mathcal{H}=
\left(
\begin{array}{cc}
\Xi & \Delta \\
 -\Delta^{*} & -\Xi^{{T}}
\end{array}
\right),
\label{eq: BdG hamiltonian}
\end{eqnarray}
where $\mathcal{H}$ is a $4N \times 4N$ matrix 
for a system with $N$ orbitals (lattice sites),
and 
$\boldsymbol{c}=
\left(\boldsymbol{c}_{\uparrow},\boldsymbol{c}_{\downarrow}\right)$
is a $2N$ component vector.
[$\boldsymbol{c}$ and $\boldsymbol{c}^{\dag}$ 
can be either column or row vector
depending on the context.]
Following the notations in Ref.\ \onlinecite{Schnyder08}, 
we use two sets of $2\times 2$ Pauli matrices 
$t_{0,x,y,z}$ and $s_{0,x,y,z}$,
which act on particle-hole and spin indices, respectively. 
Because of
\begin{eqnarray}
\Xi \!\!&=&\!\!\Xi^{\dag} \quad \mbox{(hermiticity)},
\nonumber \\
\Delta \!\!&=&\!\!-\Delta^{{T}} \quad 
\mbox{(Fermi statistics)},
\end{eqnarray}
the BdG Hamiltonian (\ref{eq: BdG hamiltonian})
satisfies particle-hole symmetry (PHS)
\begin{eqnarray}
(a):&&
\mathcal{H}=-t_{x}\mathcal{H}^{{T}} t_{x},
\quad
\mbox{(PHS)}.
\label{eq: def PHS (triplet)}
\end{eqnarray}
The presence or absence of TRS 
and $\mathrm{SU}(2)$ spin rotation symmetry 
are represented by
\begin{eqnarray}
(b):&&
\mathcal{H}= {i}s_y \mathcal{H}^{{T}} (-{i}s_y),
\quad
\mbox{(TRS)},
\end{eqnarray}
and
\begin{eqnarray}
(c):&&
\big[\mathcal{H},J_{a}\big]=0,
\quad
J_{a}:=
\left(
\begin{array}{cc}
s_{a} & 0 \\
0 & -s_{a}^{{T}}
\end{array}
\right),
\nonumber \\
&&
\quad
a=x,y,z,
\quad
\mbox{[SU(2) symmetry]},
\end{eqnarray}
respectively. 

The ensemble of BdG Hamiltonians
(\ref{eq: BdG hamiltonian}) with 
PHS condition $(a)$ defines symmetry class D
of Altland and Zirnbauer.
With additional TRS condition $(b)$, 
the resulting ensemble of BdG Hamiltonians
is called  symmetry class DIII.
For both symmetry classes, spin rotation symmetry
$(c)$ is not necessary.

The Hamiltonian in symmetry class D can be thought of as,
because of PHS $(a)$, 
a single-particle Hamiltonian of Majorana fermions.
The Majorana structure of the 
BdG Hamiltonians can be revealed by
\begin{align}
&
\left(
\begin{array}{cc}
\boldsymbol{c}\\
\boldsymbol{c}^{\dag}\\
\end{array}
\right)
\to 
\frac{1}{\sqrt{2}}
\left(
\begin{array}{cc}
\boldsymbol{\eta} \\
\boldsymbol{\chi}
\end{array}
\right)
=
\frac{1}{\sqrt{2}}
\left(
\begin{array}{cc}
\boldsymbol{c} + \boldsymbol{c}^{\dag}\\
{i}\left(\boldsymbol{c} - \boldsymbol{c}^{\dag}\right)\\
\end{array}
\right)
\end{align}
where
$\eta$ and $\chi$ are Majorana fermions satisfying 
\begin{align}
\eta_i \eta_j + \eta_j \eta_i = 2\delta_{ij},
\quad 
\eta^{\dag}=\eta,
\quad 
(i=1,\ldots, 2N), 
\quad \mbox{etc.}
\end{align}
Then, in this Majorana basis, the BdG Hamiltonian can be written as
\begin{align}
H &=
\left(
\begin{array}{cc}
\boldsymbol{\eta},& \boldsymbol{\chi}
\end{array}
\right)
\mathcal{X}
\left(
\begin{array}{c}
\boldsymbol{\eta}\\
 \boldsymbol{\chi}
\end{array}
\right)
\end{align}
where
\begin{align}
&
\mathcal{X}=
\frac{1}{2}
\left(
\begin{array}{cc}
P + S &  -{i}\left(Q - R\right) \\
{i}\left(Q + R\right) &
P - S\\
\end{array}
\right),
\end{align}
and 
\begin{align}
&
P = \Xi - \Xi^T=-P^T,\quad
Q = \Xi + \Xi^T=+Q^T,\quad
\nonumber \\
&
R = \Delta +\Delta^*=-R^T,\quad
S = \Delta -\Delta^*=+S^T.
\end{align}
Then, the $4N\times 4N$ matrix $\mathcal{X}$ satisfies
\begin{align}
\mathcal{X}^{\dag} =\mathcal{X},
\quad
\mathcal{X}^T = -\mathcal{X}.
\label{eq: class D majorana}
\end{align}
These conditions define symmetry class D.
On the other hand, symmetry class DIII is defined by, in addition, 
\begin{align}
{i}s_y \mathcal{X}^T  (-{i}s_y)
=\mathcal{X}.
\label{eq: class DIII majorana}
\end{align}

While it is always possible to cast the BdG Hamiltonians 
into a form of a single particle Hamiltonian of Majorana fermions, 
by rewriting the BdG Hamiltonian in terms of 
the "real" and "imaginary" parts of the electron operator, 
$\boldsymbol{\eta}$ and $\boldsymbol{\chi}$, 
there is no natural way in general to rewrite Majorana hopping 
problems as a BdG Hamiltonian. 
In order to do so, 
the single particle Majorana Hamiltonian must be 
an even-dimensional matrix, and 
we need to specify a particular 
way to make a complex fermion operator out 
of two Majorana fermion operator. 
Still, any single-particle Hamiltonian for Majorana fermions,
with its defining properties (\ref{eq: class D majorana}),
can be classified in terms of the presence (class DIII)
or absence (class D) of TRS (\ref{eq: class DIII majorana})
without referring to complex fermions.

\paragraph{off-diagonal block structure of class DIII Hamiltonians}
Combining class DIII conditions
$(a)$ and $(b)$, one can see that
a member of class DIII anticommutes 
with a unitary matrix
$t_xs_y$,
\begin{eqnarray}
\mathcal{H} 
\!\!&=&\!\!
 -t_x s_y \mathcal{H} s_y t_x.
\end{eqnarray}
In this sense, 
class DIII Hamiltonians can be said to have a chiral structure.
In order to compute the winding number $\nu$,
defined for class DIII Hamiltonians in three spatial dimensions,
it is necessary to go to a
basis in which the chiral transformation
$t_x s_y$ is diagonal.
We can find such a basis as follows:
we first rotate
$t_x \to t_z$ and
$s_y \to s_z$ by
a unitary transformation
\begin{eqnarray}
W_1 \!\!&=&\!\!
\frac{1}{\sqrt{2}}
\left(t_0+{i}t_y\right)
\frac{1}{\sqrt{2}}
\left(s_0-{i}s_x\right),
\end{eqnarray}
I.e., $W_1 t_x W^{\dag}_1
=
-t_z,$
and
$W_1 s_y W^{\dag}_1
=
-s_z$.
We then exchange the 3rd and 4th entries by
a unitary transformation $W_2$, 
\begin{eqnarray}
W_2 W_1 t_x s_y W^{\dag}_1W^{\dag}_2
\!\!&=&\!\!
W_2 t_z s_z W^{\dag}_2
=
t_0 s_z.
\end{eqnarray}
Further exchanging the 2nd and 3rd entries
by a unitary transformation $W_3$, 
the combined unitary transformation 
$W=W_3 W_2 W_1$ diagonalizes $t_x s_y$,
\begin{eqnarray}
t_x s_y
\!\!&\to &\!\!
W t_x s_y W^{\dag}
=
t_z s_0.
\end{eqnarray}
Under the transformation $W$
PHS and TRS transformations
are transformed as
\begin{eqnarray}
t_x s_0
\!\!&\to &\!\!
W t_x s_0 W^T
=
-{i}t_x s_z,
\quad
\mbox{(PHS)}
\nonumber \\
t_0 s_y
\!\!&\to &\!\!
W t_0 s_y W^T
=
t_y s_z, 
\quad
\mbox{(TRS)}
\end{eqnarray}
respectively. 
Observe the transformed PHS and TRS
pick up the same sign under matrix transposition 
[$
(t_x s_z)^T=
+(t_x s_z)^T
$
and
$(t_y s_z)^T=
-(t_y s_z)$,
respectively]
as the original ones
[$(t_x s_0)^T=
+(t_x s_0)$
and
$
(t_0 s_y)^T=
-(t_0 s_y)^T$,
respectively].
In this basis, the Hamiltonian takes
the block-off diagonal form,
\begin{eqnarray}
\mathcal{H}
&\to &
\left(
\begin{array}{cc}
0 & D \\
D^{\dag} & 0
\end{array}
\right),
\quad
D= -s_z D^T s_z.
\end{eqnarray}
This can be further simplified by a unitary transformation
\begin{eqnarray}
\mathcal{H}
\!\!&\to &\!\!
\left(
\begin{array}{cc}
0 & s^{\dag}_{xy} \\
s_{xy} & 0
\end{array}
\right)
\left(
\begin{array}{cc}
0 & D \\
D^{\dag} & 0
\end{array}
\right)
\left(
\begin{array}{cc}
0 & s^{\dag}_{xy} \\
s_{xy} & 0
\end{array}
\right)
\nonumber \\
 \!\!&=&\!\!
 \left(
 \begin{array}{cc}
 0 & s^{\dag}_{xy}D^{\dag}s^{\dag}_{xy} \\
 s_{xy} D s_{xy} & 0
 \end{array}
 \right),
 \end{eqnarray}
where
\begin{align}
s_z = -{i} s^T_{xy} s^{\ }_{xy},
\quad
s^T_{xy} = 
\frac{1}{\sqrt{2}}\left(s_x -s_y\right).
\end{align}
Introducing 
\begin{eqnarray}
D' \!\!&:=&\!\! 
s_{xy} D s_{xy} = 
- s^T_{xy} D^T s^T_{xy} = -(D')^T,
\end{eqnarray}
we finally arrive at
\begin{align}
\mathcal{H}
&\to 
\left(
\begin{array}{cc}
0 & D' \\
D^{\prime \dag} & 0
\end{array}
\right),
\quad
D'= -D^{\prime T}.
\end{align}



\begin{thebibliography}{99}
\bibitem{footnote}
Insulators and superconductors are seemingly very different states 
 of matter. In Refs.\, \onlinecite{Schnyder08} and \onlinecite{KitaevKtheory},
by focusing on the fermionic band structures
 of Bogoliubov-de Gennes (BdG) quasiparticles in superconductors 
while treating the bosonic degrees
 of freedom (condensate) in superconductors at mean field level, 
 topological nature of band structure in insulators and BdG Hamiltonians 
 are classified in a unified fashion.
\end{thebibliography}
\end{document}